\documentclass[aps,pre,twocolumn,superscriptaddress,floatfix]{revtex4-2}

\usepackage{graphicx}
\usepackage{amsmath,empheq}
\usepackage{amssymb}
\usepackage{ifthen}
\usepackage{booktabs}

\usepackage{graphicx}
\usepackage{dcolumn}
\usepackage{bm}
\usepackage{gensymb}

\newcommand{\kB}{k_{\rm B}}

\newcommand{\SM}{Supplemental Material}

\newcommand{\bb}{\begin{equation}}
\newcommand{\ee}{\end{equation}}
\newcommand{\ba}{\begin{eqnarray*}}
\newcommand{\ea}{\end{eqnarray*}}

\makeatletter
\def\@frameeq#1{%
  \framebox{$\,\displaystyle#1\hbox{\vrule height 2.4ex depth 1.4ex width 0pt}\,$}}
\newcommand\Equation[1]{$$\refstepcounter{equation}%
  \@frameeq{#1}%
  \eqno \hbox{\@eqnnum}$$\@ignoretrue\ignorespaces}
\newcommand\Displaystyle[1]{$$\@frameeq{#1}$$\@ignoretrue\ignorespaces}
\makeatother

\usepackage{longtable}
\usepackage[caption=false]{subfig}
\usepackage[allcolors=black,colorlinks]{hyperref}
 \usepackage{nameref}

\begin{document}

\title{First-order condensation and meniscus pinning in finite open wedges}

\author{Ji\v{r}\'i Janek}
\affiliation{Research group of Molecular and Mesoscopic Modelling, The Czech Academy of Sciences, Institute of Chemical Process Fundamentals, 165 02 Prague, Czech Republic}

\author{Miriam Mago\v{c}iov\'a}
\affiliation{Department of Physical Chemistry, University of Chemistry and Technology Prague, 166 28 Prague, Czech Republic}

\author{Alexandr Malijevsk\'y}
\affiliation{Research group of Molecular and Mesoscopic Modelling, The Czech Academy of Sciences, Institute of Chemical Process Fundamentals, 165 02 Prague, Czech Republic}
\affiliation{Department of Physical Chemistry, University of Chemistry and Technology Prague, 166 28 Prague, Czech Republic}

\begin{abstract}
We study first-order condensation in finite open wedges using classical density functional theory (DFT) and compare the resulting microscopic phase behaviour with recent macroscopic predictions. Two geometries are considered: a symmetric wedge formed by two identical inclined walls and an asymmetric wedge formed by a vertical and an inclined wall. The DFT calculations confirm the two condensed morphologies predicted for the symmetric wedge and the three morphologies predicted for the asymmetric wedge and their emergence depending on the wedge size, opening angle, and wall contact angle. Under partial-wetting conditions, the microscopic phase boundaries are in nearly perfect quantitative agreement with the macroscopic predictions, including reentrant changes between pinned and depinned meniscus configurations and the appearance of the semi-double-pinned state in the asymmetric wedge. Under complete-wetting conditions, the DFT phase diagrams retain the topology predicted by the macroscopic theory, but the single-pinned region is systematically enlarged. This shift is attributed to adsorbed liquid films, which reduce the effective width available to the gas and thereby increase the effective aspect ratio of the confinement.
\end{abstract}

\maketitle

\section{Introduction}

Fluid phase behaviour under confinement is governed not only by the
surface properties of the confining walls but also by their geometry
and spatial extent. Even when the underlying bulk fluid exhibits only a
conventional liquid-gas transition, confinement may shift phase
coexistence, stabilize new interfacial morphologies, and induce additional phase transitions, see e.g. Refs.~ \cite{Cahn1977, Fisher1981, Evans1990, Gelb2000, Rowlinson, Binder2008, Herminghaus2008, Quere2008, Henderson2009, Bonn2009, Saam_2009, Cardenas2020}. 



A paradigmatic example  of geometry-induced interfacial
behaviour is provided by wedge-shaped confinement \cite{Concus1969, Hauge1992, Parry1999, Rejmer1999, Bruschi2002, Abraham2002, Abraham2003, Hernandez2006, PRL2013, Malijevsky2013, Delfino_2014, Malijevsky2015, Klomkliang2015, Loi_2020, Turci_2024}. Consider an infinite
linear wedge formed by two planar walls with opening angle $\phi$.  At bulk liquid-gas coexistence, macroscopic theory predicts a filling transition at the temperature $T_f$ satisfying\cite{Concus1969, Hauge1992}
\begin{equation}
	\theta(T_f)=\frac{\pi-\phi}{2},
	\label{eq:wedge-filling}
\end{equation}
where $\theta(T)$ is Young's contact angle corresponding to the confining walls. This relation provides a direct connection between the thermodynamic wetting properties of the wall and the geometry of the confinement.

For $T<T_f$, adsorption near the wedge apex remains microscopic, whereas for $T>T_f$ the wedge is macroscopically filled as bulk coexistence is approached. Since the contact angle typically decreases with increasing temperature, filling generally precedes planar wetting, $T_f<T_w$, where $T_w$ denotes the wetting temperature at which $\theta$ vanishes~\cite{Dietrich1988, Schick}. A wedge may therefore be macroscopically filled even though each of its walls remains only partially wet.

The filling transition at coexistence may
be either first order or continuous (critical) \cite{Parry1999, Rejmer1999, Malijevsky2013}; in the latter case, the
equilibrium height of the meniscus formed above the wedge apex diverges
continuously as $T\to T_f^{-}$, whilst  first-order filling is characterized by a coexistence between a partially filled and completely filled wedge. For $T>T_f$, complete filling occurs on
approaching bulk coexistence from the gas side, with the meniscus
flattening and receding indefinitely from the wedge apex. Microscopic density-functional
studies have confirmed both the thermodynamic filling condition and the
occurrence of first-order and critical filling
transitions depending on the microscopic wall-fluid interactions and the wedge opening angle~\cite{PRL2013, Malijevsky2013, Malijevsky2015}.

The finite open wedges studied here differ fundamentally from their
infinite counterparts. Because the walls terminate at both ends, the
condensed liquid is bounded by two menisci and condensation occurs
through a first-order jump at a pressure below bulk saturation, $p<p_{\rm sat}$. The presence of wall edges also allows one or both menisci to become pinned, producing distinct condensation morphologies that have no
analogue in an infinite wedge.

In our previous work, we first considered a symmetric truncated wedge
formed by two identical finite walls inclined symmetrically with respect
to the wedge axis~\cite{Janek2024}. The wedge has  narrow
and wide openings, and the condensed phase is bounded by two menisci.
The meniscus at the narrow opening is always pinned, whereas the second
meniscus may either lie inside the wedge or become pinned at the
wider opening. These two possibilities define the single-pinned (SP)
and double-pinned (DP) condensation states, respectively. A
macroscopic free-energy analysis yielded Kelvin-like equations for both
regimes, the conditions for their stability, and the corresponding
phase diagram. In particular, the theory predicts a reentrant sequence
$\mathrm{DP}\rightarrow\mathrm{SP}\rightarrow\mathrm{DP}$
upon varying  the wall inclination for appropriate wall lengths and
contact angles. The SP-DP transition is continuous and corresponds to the (de)pinning of the meniscus at the wider opening.

We subsequently considered an asymmetric open wedge formed by a vertical wall and an inclined wall with the same vertical extent~\cite{Janek2025}. Breaking the symmetry separates the pinning of the wide-end meniscus at the two wall edges and gives rise to an intermediate morphology. In addition to the SP and DP states, the system may exhibit a semi-double-pinned (SDP) state, in which the wide-end meniscus is pinned at the upper edge of the vertical wall while meeting the inclined wall at Young's contact angle. The macroscopic theory predicts several classes of phase diagrams depending on the wedge aspect ratio and identifies the conditions under which the three condensation morphologies occur.  In particular, while condensation (of any type) is still constrained by the corresponding infinite-wedge filling condition given by Eq.~(1) (with $\phi=\alpha$), the DP configuration requires an additional constraint $\alpha+\theta<\pi/2$. 

Both theories are based on a sharp-interface macroscopic description in which the gas and liquid are homogeneous bulk phases separated by circular menisci, while the wall adsorption properties enter only through Young's contact angle and the corresponding surface free energies. This construction provides a transparent geometrical interpretation of the condensation mechanisms and yields explicit or implicit predictions for the phase boundaries. At the same time, it neglects molecular-scale effects such as density oscillations near the walls, diffuse liquid--gas interfaces, adsorption films, and microscopic structure near edges and corners. These effects are expected to be particularly important under complete-wetting conditions, where liquid films form on the walls before condensation and reduce the effective space available to the gas.

The purpose of the present work is to test these macroscopic predictions using microscopic classical density functional theory (DFT). We study the same symmetric and asymmetric open wedges and determine both the condensation point and the resulting morphology from the equilibrium density profiles. Calculations are performed for partially wet walls and under complete-wetting conditions, allowing us to assess separately the validity of the sharp-interface description when adsorption remains microscopic and when substantial liquid films are present.

The condensed states are identified using complementary density-based and geometrical criteria, described in detail in the Supplemental Material. The resulting microscopic phase diagrams are then compared directly with the macroscopic predictions. For partially wet walls, we find nearly perfect quantitative agreement, including the onset of SP condensation, the appearance of the SDP state in the asymmetric wedge, and the predicted reentrant condensation changes. For completely wet walls, the overall
topology of the phase diagrams remains correct, but systematic shifts
of the phase boundaries occur.  These deviations can be  attributed  to adsorbed liquid films, which effectively narrow the confinement and favour morphologies characteristic of larger aspect ratios.

The remainder of the paper is organized as follows. Section~II summarizes the macroscopic theories for the symmetric and asymmetric wedges. Section~III introduces the microscopic fluid model, density functional, wall potentials, and numerical procedure. Section~IV presents the DFT results and their comparison with the macroscopic phase diagrams. Finally, Sec.~V summarizes the main conclusions. Details of the classification of the condensed states and the derivation of the external wall potentials are provided in the Supplemental Material.

\section{Macroscopic theory}
\label{sec:theory}

In this section, we briefly summarize the macroscopic theories of
condensation in the two open-wedge geometries considered below
\cite{Janek2024,Janek2025}. In both cases, the
system is translationally invariant in the direction perpendicular to
the wedge cross section and is in contact with a bulk gas reservoir at
a subcritical temperature $T$ and pressure $p$.

At the macroscopic level, a condensed state consists of a region of
metastable liquid bounded by two circular menisci separating it from
the surrounding gas. Since both menisci are subject to the same
pressure difference, they have a common Laplace radius
\begin{equation}
	R=\frac{\gamma}{\delta p},
	\label{eq:laplace_radius}
\end{equation}
where $\gamma$ is the liquid-gas surface tension and
$\delta p=p-p_l$ is the pressure difference between the reservoir gas
and the metastable liquid stabilized inside the wedge. In general, a meniscus may either lie inside the wedge or be attached to one or both wall edges. In the former case, the meniscus meets the walls  at the equilibrium Young contact angle $\theta$ and is free to slide along them upon changing the pressure. By contrast, a meniscus attached to a wall edge is geometrically pinned
and need not satisfy Young's condition there. Instead, it meets the wall at the {\emph edge contact angle} $\theta_e$, which is geometry-dependent \cite{Malijevsky2017, Malijevsky2021a}.

The excess grand potential of a given condensed configuration contains
a positive bulk contribution associated with the metastable liquid,
the interfacial free-energy cost of the two menisci, and a negative wall
contribution arising from the replacement of wall-gas interfaces by
wall-liquid interfaces. The condensation pressure is determined by
the condition that this excess grand potential vanishes, so that the
condensed and gas-like states coexist.

\subsection{Symmetric open wedge}
\label{subsec:symmetric_theory}

\begin{figure}
	\centering%
	\subfloat[Single-pinned state.\label{subfig:sym-SP-scheme}]{%
		\includegraphics[width=0.7\columnwidth]{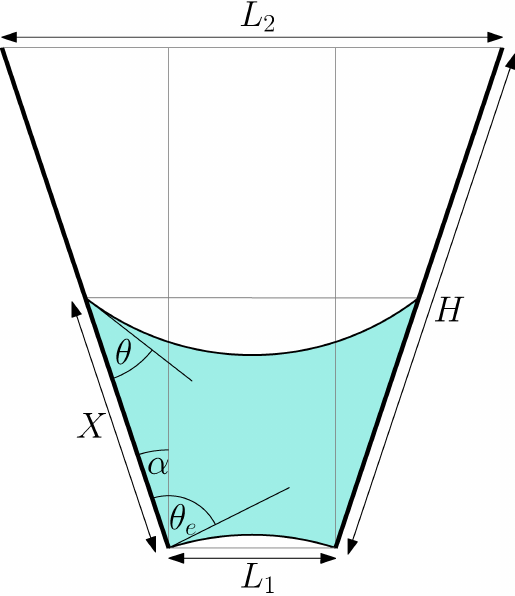}%
	}%
	\\%
	\subfloat[Double-pinned state.\label{subfig:sym-DP-scheme}]{%
		\includegraphics[width=0.7\columnwidth]{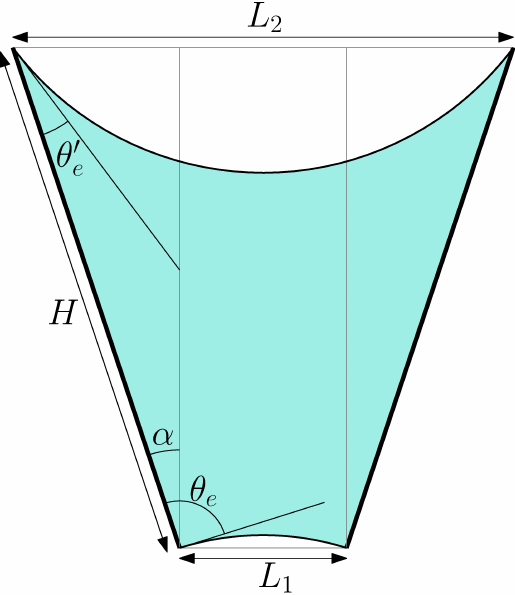}%
	}%
\caption{Schematic illustration of the symmetric open wedge formed by two identical walls of length $H$, each inclined by an angle $\alpha$ with respect to the symmetry axis. The widths of the narrow and wide openings are $L_1$ and $L_2$, respectively, with $L_2=L_1+2H\sin\alpha$. The condensed liquid is bounded by two circular menisci of common Laplace radius $R$. (a) In the single-pinned state, the narrow-end meniscus is pinned and characterized by the edge contact angle $\theta_e$, whereas the second meniscus lies inside the wedge at a distance $X$ from the narrow opening and meets both walls at Young's contact angle $\theta$. (b) In the double-pinned state, both menisci are pinned at the wedge openings and are characterized by the edge contact angles $\theta_e$ and $\theta'_e$ at the narrow and wide ends, respectively.} 	\label{fig:symmetric_geometry}
\end{figure}

%
%
%

The symmetric geometry consists of two identical walls of length $H$,
each inclined by an angle $\alpha$ relative to the symmetry axis, as
illustrated in Fig.~\ref{fig:symmetric_geometry}. The widths of the
narrow and wide openings are $L_1$ and
\begin{equation}
	L_2=L_1+2H\sin\alpha,
	\label{eq:sym_L2}
\end{equation}
respectively.

In a condensed state, the meniscus at the narrow opening is always pinned. 
The position of the second meniscus determines whether the condensed state is single-pinned or double-pinned:

\begin{enumerate}

\item In the {\bf single-pinned (SP)} state,  the meniscus located  at the narrow end meets the walls at the edge contact angle $\theta_e$, while the other meniscus lies inside the wedge and thus meets both walls at Young's angle $\theta$; see Fig.~\ref{fig:symmetric_geometry}(a). The pressure at which SP condensation occurs is determined by the Kelvin-like equation
\begin{equation}
	\delta p_{\rm SP}=\frac{2\gamma\cos(\theta_e-\alpha)}{L_1},
	\label{eq:sym_SP_kelvin}
\end{equation}
where the value of $\theta_e$ at condensation follows from the coexistence condition and is determined implicitly by
\begin{align}
	&\cot\alpha\left[\cos^2(\alpha+\theta)
	-\cos^2(\alpha-\theta_e)\right]+\pi-\theta-\theta_e
	\nonumber\\
	&+\frac{\sin(2\alpha+2\theta)-\sin(2\alpha-2\theta_e)}{2}
	\nonumber\\
	&=\frac{2\cos\theta\left[\cos(\alpha+\theta)
		-\cos(\alpha-\theta_e)\right]}{\sin\alpha}.
	\label{eq:sym_SP_thetae}
\end{align}

For a given pressure difference, and hence a given Laplace radius $R$, the position of the free meniscus is characterized by its distance $X$ from the narrow opening, measured along either wall:
\begin{equation}
	X=\frac{2R\cos(\alpha+\theta)-L_1}{2\sin\alpha}.
	\label{eq:sym_X}
\end{equation}
The SP configuration is admissible only while the free meniscus remains inside the wedge, $X<H$.

\item In the {\bf double-pinned (DP)} state, both menisci are pinned at the wedge openings; see Fig.~\ref{fig:symmetric_geometry}(b).

Having the same radius, the narrow- and wide-end menisci are characterized by different edge contact angles, $\theta_e$ and $\theta'_e$, respectively, satisfying
\begin{equation}
	R=\frac{L_1}{2\cos(\theta_e-\alpha)}
	=\frac{L_2}{2\cos(\theta'_e+\alpha)}.
	\label{eq:sym_DP_geometry}
\end{equation}
The free-energy balance implies the condition
\begin{equation}
	\begin{aligned}
		&\frac{L_2^2-L_1^2}{4}\cot\alpha+R^2(\pi-\theta_e-\theta'_e)
		-\frac{(L_2-L_1)R\cos\theta}{\sin\alpha} \\
		&+\frac{R}{2}\left[L_1\sin(\theta_e-\alpha)
		+L_2\sin(\theta'_e+\alpha)\right]=0.
	\end{aligned}
	\label{eq:sym_DP_coexistence}
\end{equation}
Together, Eqs.~\eqref{eq:sym_DP_geometry} and
\eqref{eq:sym_DP_coexistence} determine $R$, $\theta_e$, and
$\theta'_e$, and hence the pressure difference
\begin{equation}
	\delta p_{\rm DP}=\frac{\gamma}{R},
	\label{eq:sym_DP_kelvin}
\end{equation}
at which DP condensation occurs.

\end{enumerate}

The boundary between SP and DP condensation is reached when the free
meniscus in the SP state arrives at the wide opening,
\begin{equation}
	X=H, \qquad \text{(SP-DP boundary)},
	\label{eq:sym_boundary_X}
\end{equation}
or, equivalently, when
\begin{equation}
	\theta'_e=\theta, \qquad \text{(SP-DP boundary)}.
	\label{eq:sym_boundary_angle}
\end{equation}
Because $X$ depends non-monotonically on $\alpha$, this condition may
have two solutions, leading to the reentrant sequence
${\rm DP}\rightarrow{\rm SP}\rightarrow{\rm DP}$.

Capillary condensation is possible only for sufficiently long walls,
\begin{equation}
	H>L_1\sec\theta,
	\label{eq:sym_H_cond}
\end{equation}
and for wall inclinations not exceeding $\alpha_{\max}$, determined by
\begin{equation}
	\sin\alpha_{\max}=\cos\theta-\frac{L_1}{H}.
	\label{eq:sym_alpha_max}
\end{equation}
Equation~\eqref{eq:sym_alpha_max} is the finite-length extension of the infinite-wedge filling criterion, Eq.~\eqref{eq:wedge-filling}. Indeed, in the limit $H\to\infty$, it reduces to $\sin\alpha_{\max}=\cos\theta$, or $\theta=\pi/2-\alpha_{\max}$. Since the full opening angle of the symmetric wedge is $\phi=2\alpha$, this is precisely Eq.~\eqref{eq:wedge-filling}. For a finite wedge, the condensation criterion is therefore more restrictive.

The SP state requires still longer walls, and the corresponding minimum length $H_c$ depends on $\theta$. Its global minimum is $H_c\simeq4.5L_1$, attained at $\theta\simeq11.6^\circ$.

\begin{figure*}[t]
	\centering
	\begin{minipage}[t]{0.32\textwidth}
		\centering
		\includegraphics[height=18em,width=\linewidth,keepaspectratio]{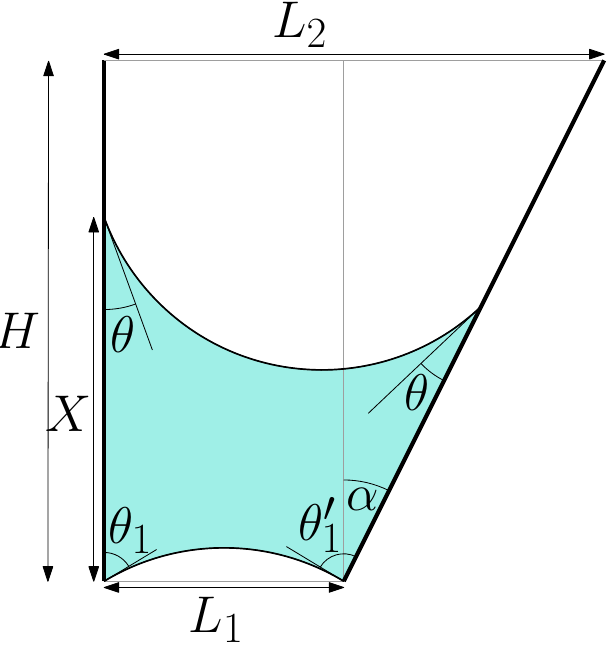}
		
		\smallskip
		(a) SP
	\end{minipage}
	\hfill
	\begin{minipage}[t]{0.32\textwidth}
		\centering
		\includegraphics[height=18em,width=\linewidth,keepaspectratio]{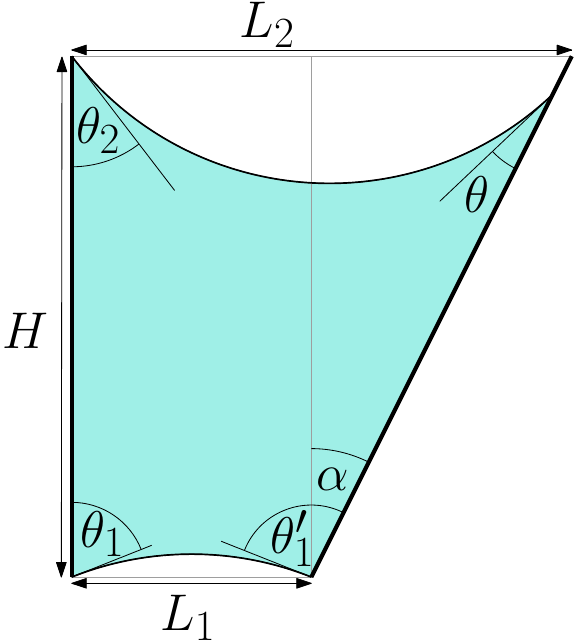}
		
		\smallskip
		(b) SDP
	\end{minipage}
	\hfill
	\begin{minipage}[t]{0.32\textwidth}
		\centering
		\includegraphics[height=18em,width=\linewidth,keepaspectratio]{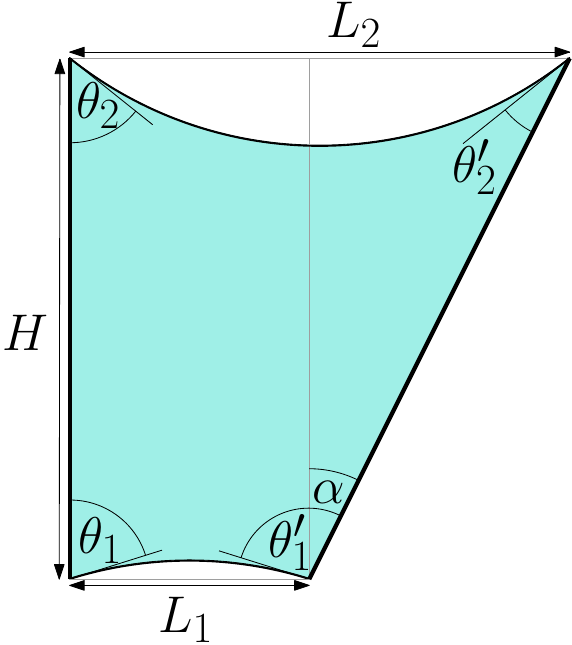}
		
		\smallskip
		(c) DP
	\end{minipage}
\caption{Schematic illustration of the asymmetric open wedge formed by a vertical wall and an inclined wall whose vertical projections both have length $H$. The angle between the walls is $\alpha$, and the widths of the narrow and wide openings are $L_1$ and $L_2$, respectively, with $L_2=L_1+H\tan\alpha$. In all cases, the condensed liquid is bounded by two circular menisci of common Laplace radius $R$, with the lower meniscus pinned at the narrow opening. (a) Single-pinned (SP) state: the upper meniscus lies inside the wedge and meets both walls at Young's contact angle $\theta$. (b) Semi-double-pinned (SDP) state: the upper meniscus is pinned at the upper edge of the vertical wall while meeting the inclined wall at Young's contact angle. (c) Double-pinned (DP) state: the upper meniscus is pinned at both edges of the wide opening.}
	\label{fig:asymmetric_geometry}
\end{figure*}

\subsection{Asymmetric open wedge}
\label{subsec:asymmetric_theory}

The asymmetric wedge model consists of a vertical wall and an inclined wall whose vertical projections both have length $H$, as illustrated in Fig.~\ref{fig:asymmetric_geometry}. The angle between the walls is $\alpha$, and the widths of the narrow and wide openings are related by
\begin{equation}
	L_2=L_1+H\tan\alpha.
	\label{eq:asym_L2}
\end{equation}

In this model, condensation may lead to  three distinct  states: single-pinned (SP), semi-double-pinned (SDP), or double-pinned (DP); see Fig.~\ref{fig:asymmetric_geometry}. In all three states, the lower meniscus is pinned at the narrow opening. If $\theta_1$ denotes its edge contact angle with the vertical wall, its angle with the inclined wall is then $\theta'_1=\theta_1+\alpha$ and its geometry gives
\begin{equation}
	\cos\theta_1=\frac{L_1}{2R}.
	\label{eq:asym_lower_geometry}
\end{equation}
Thus all three condensation regimes obey the common Kelvin-like relation
\begin{equation}
	\delta p_i=\frac{2\gamma\cos\theta_1}{L_1}, \qquad i={\rm SP,SDP,DP},
	\label{eq:asym_kelvin}
\end{equation}
with $\theta_1$ determined by the geometry and free-energy balance specific to each morphology.

\begin{enumerate}
	
\item In the \textbf{single-pinned (SP)} state, the upper meniscus lies inside the wedge and meets both walls at Young's contact angle $\theta$; see Fig.~\ref{fig:asymmetric_geometry}(a). For a given pressure, and hence a given Laplace radius $R$, its position is characterized by the distance $X$ of its contact point on the vertical wall from the narrow opening:
\begin{equation}
	\begin{aligned}
		X={}&R\cot\alpha\left[\cos\theta+\cos(\theta+\alpha)\right] \\
		&+R\left[\sin(\theta+\alpha)-\sin\theta\right]-L_1\cot\alpha.
	\end{aligned}
	\label{eq:asym_SP_X}
\end{equation}
The free-energy balance yields 
\begin{equation}
	\begin{aligned}
		&\frac{1}{2}(\cot\alpha+\csc\alpha)\left(\cos^2\theta-2\cos\theta_1\cos\theta\right) \\
		&-\frac{1}{2}\left(\sin\theta_1\cos\theta_1+\sin\theta\cos\theta\right)+\cos^2\theta_1\cot\alpha \\
		&-\frac{\pi}{2}+\frac{\alpha}{4}+\frac{\theta+\theta_1}{2}=0,
	\end{aligned}
	\label{eq:asym_SP_condition}
\end{equation}
which determines $\theta_1$ and hence the SP condensation pressure. The free-energy condition is independent of the wall length $H$, provided that $X<H$, reflecting the fact that the position and shape of an internal meniscus are unaffected by extending the walls. 

	\item In the \textbf{semi-double-pinned (SDP)} state, the upper meniscus is pinned at the upper edge of the vertical wall while meeting the inclined wall at Young's contact angle $\theta$; see Fig.~\ref{fig:asymmetric_geometry}(b). Denoting its edge contact angle with the vertical wall by $\theta_2$, its geometry requires
	\begin{equation}
		\frac{L_2}{R}=\frac{\cos\theta+\cos(\theta_2+\alpha)}{\cos\alpha},
		\label{eq:asym_SDP_geometry}
	\end{equation}
	while  the free-energy balance implies that
	\begin{equation}
		\begin{aligned}
			&(\cot\alpha+2\csc\alpha)\cos^2\theta-\sin\theta\cos\theta-\sin(2\theta_1)+\sin\theta_2\cos\theta_2 \\
			&+\left[2\cot\alpha(\cos\theta_2-2\cos\theta_1)-4\cos\theta_1\csc\alpha-2\sin\theta_2\right]\cos\theta \\
			&+\left(4\cos^2\theta_1-\cos^2\theta_2\right)\cot\alpha=2\pi-\alpha-\theta-2\theta_1-\theta_2.
		\end{aligned}
		\label{eq:asym_SDP_condition}
	\end{equation}
	Equations~\eqref{eq:asym_lower_geometry}, \eqref{eq:asym_SDP_geometry}, and \eqref{eq:asym_SDP_condition} determine $\theta_1$, $\theta_2$, and $R$, and hence the condensation pressure $\delta p_{\rm SDP}$.
	
	\item In the \textbf{double-pinned (DP)} state, the upper meniscus is pinned at both edges of the wide opening; see Fig.~\ref{fig:asymmetric_geometry}(c). Let $\theta_2$ denote its edge contact angle with the vertical wall. Its contact angle with the inclined wall is then $\theta'_2=\theta_2-\alpha$, and its geometry implies
	\begin{equation}
		\cos\theta_2=\frac{L_2}{2R}.
		\label{eq:asym_DP_geometry}
	\end{equation}
	The free-energy balance gives
	\begin{equation}
		\begin{aligned}
			&H(1+\sec\alpha)\cos\theta
			-R\sin(\theta_1+\theta_2)\cos(\theta_1-\theta_2)\\
			&=(\cos\theta_1+\cos\theta_2)H+R(\pi-\theta_1-\theta_2)
		\end{aligned}
		\label{eq:asym_DP_condition}
	\end{equation}
	Together, Eqs.~\eqref{eq:asym_lower_geometry}, \eqref{eq:asym_DP_geometry}, and \eqref{eq:asym_DP_condition} determine $\theta_1$, $\theta_2$, and $R$, and hence the pressure difference $\delta p_{\rm DP}$.
	
\end{enumerate}

%
%
	

\section{Microscopic model}
\label{sec:dft}

We consider a one-component fluid confined by the open-wedge geometries introduced in Sec.~II and in equilibrium with a bulk reservoir at temperature $T$ and chemical potential $\mu$. Within classical density functional theory (DFT) \cite{Evans1979}, the equilibrium one-body density $\rho(\mathbf r)$ is obtained by minimizing the grand-potential functional
\begin{equation}
	\Omega[\rho]=\mathcal F[\rho]+\int d\mathbf r\,\rho(\mathbf r)\left[V(\mathbf r)-\mu\right],
	\label{eq:grand_potential}
\end{equation}
where $\mathcal F[\rho]$ is the intrinsic Helmholtz free-energy functional and $V(\mathbf r)$ is the external potential generated by the confining walls. For a simple fluid, we employ the standard perturbative decomposition
\begin{equation}
	\mathcal F[\rho]=\mathcal F_{\rm id}[\rho]+\mathcal F_{\rm hs}[\rho]+\mathcal F_{\rm att}[\rho],
	\label{eq:free_energy_decomposition}
\end{equation}
where the three terms describe the ideal-gas contribution, short-ranged repulsive correlations, and attractive intermolecular interactions, respectively. The ideal-gas contribution is known exactly:
\begin{equation}
	\beta\mathcal F_{\rm id}[\rho]=\int d\mathbf r\,\rho(\mathbf r)\left\{\ln\!\left[\Lambda^3\rho(\mathbf r)\right]-1\right\},
	\label{eq:ideal_functional}
\end{equation}
where $\beta=(k_BT)^{-1}$ and $\Lambda$ is the thermal de Broglie wavelength, which can be set to unity.

The repulsive interactions are represented by a hard-sphere reference system and treated using the original Rosenfeld fundamental-measure theory (FMT)~\cite{Rosenfeld1989}. The corresponding excess free-energy contribution is
\begin{equation}
	\mathcal F_{\rm hs}[\rho]=\beta^{-1}\int d\mathbf r\,\Phi(\{n_\nu\}),
\end{equation}
where $\Phi$ is a local function of six weighted densities,
\begin{equation}
	n_\nu(\mathbf r)=\int d\mathbf r'\,\rho(\mathbf r')\omega_\nu(\mathbf r-\mathbf r'),
	\qquad \nu=0,1,2,3,\mathbf 1,\mathbf 2.
\end{equation}
The set comprises the four scalar weighted densities $n_0,n_1,n_2,n_3$ and the two vector densities $\mathbf n_1,\mathbf n_2$. For hard spheres of radius ${\cal R}=\sigma/2$, the scalar weight functions are
\begin{equation}
	\begin{aligned}
		\omega_3(\mathbf r)&=\Theta({\cal R}-r),&
		\omega_2(\mathbf r)&=\delta({\cal R}-r),\\
		\omega_1(\mathbf r)&=\frac{\omega_2(\mathbf r)}{4\pi{\cal R}},&
		\omega_0(\mathbf r)&=\frac{\omega_2(\mathbf r)}{4\pi{\cal R}^2}.
	\end{aligned}
\end{equation}
while the vector weight functions are
\begin{equation}
	\boldsymbol{\omega}_2(\mathbf r)=\frac{\mathbf r}{r}\delta({\cal R}-r),
	\qquad
	\boldsymbol{\omega}_1(\mathbf r)=\frac{\boldsymbol{\omega}_2(\mathbf r)}{4\pi{\cal R}}.
\end{equation}
Here $r=|\mathbf r|$, $\Theta$ is the Heaviside function, and $\delta$ is the Dirac delta distribution. Among the available prescriptions for the free-energy density $\Phi$, we use the original Rosenfeld approximation
\begin{equation}
	\Phi=
	-n_0\ln(1-n_3)
	+\frac{n_1n_2-\mathbf n_1\cdot\mathbf n_2}{1-n_3}
	+\frac{n_2^3-3n_2\,\mathbf n_2\cdot\mathbf n_2}
	{24\pi(1-n_3)^2},
\end{equation}
which in the homogeneous limit reproduces the Percus-Yevick compressibility equation of state.

For separations larger than the hard-core diameter $\sigma$, the fluid particles interact through the attractive part of a truncated Lennard--Jones potential,
\begin{equation}
	u_{\rm att}(r)=
	\begin{cases}
		0, & r<\sigma,\\
		-4\varepsilon(\sigma/r)^6, & \sigma<r<r_c,\\
		0, & r>r_c,
	\end{cases}
	\label{eq:attractive_pair_potential}
\end{equation}
$r_c=2.5\sigma$ is the interaction cutoff and $\varepsilon$ and $\sigma$ set the energy and length scales, respectively. The corresponding attractive contribution is included in \eqref{eq:free_energy_decomposition} at mean-field level
\begin{equation}
	\mathcal F_{\rm att}[\rho]=\frac{1}{2}\iint d\mathbf r\,d\mathbf r'\,
	\rho(\mathbf r)\rho(\mathbf r')u_{\rm att}(|\mathbf r-\mathbf r'|).
	\label{eq:attractive_functional}
\end{equation}

The confining walls are modelled as continuous distributions of particles of uniform number density $\rho_w$. A wall particle interacts with a fluid particle through the full Lennard-Jones potential
\begin{equation}
	\phi_w(r)=4\varepsilon_w\left[\left(\frac{\sigma_w}{r}\right)^{12}
	-\left(\frac{\sigma_w}{r}\right)^6\right],
	\label{eq:wall_pair_potential}
\end{equation}
so that the external potential is given  by integrating $\phi_w(r)$ over the entire domain  ${\mathcal W}$ of the confining walls
\begin{equation}
	V(\mathbf r)=\rho_w\int_{\mathcal W}d\mathbf r'\,
	\phi_w(|\mathbf r-\mathbf r'|),
	\label{eq:wall_field}
\end{equation}
with $\rho_w\sigma^3=1$, $\sigma_w=\sigma$, and $\varepsilon_w=\varepsilon$. 


\begin{figure}[tb]
	\centering
	\includegraphics[width=\columnwidth]{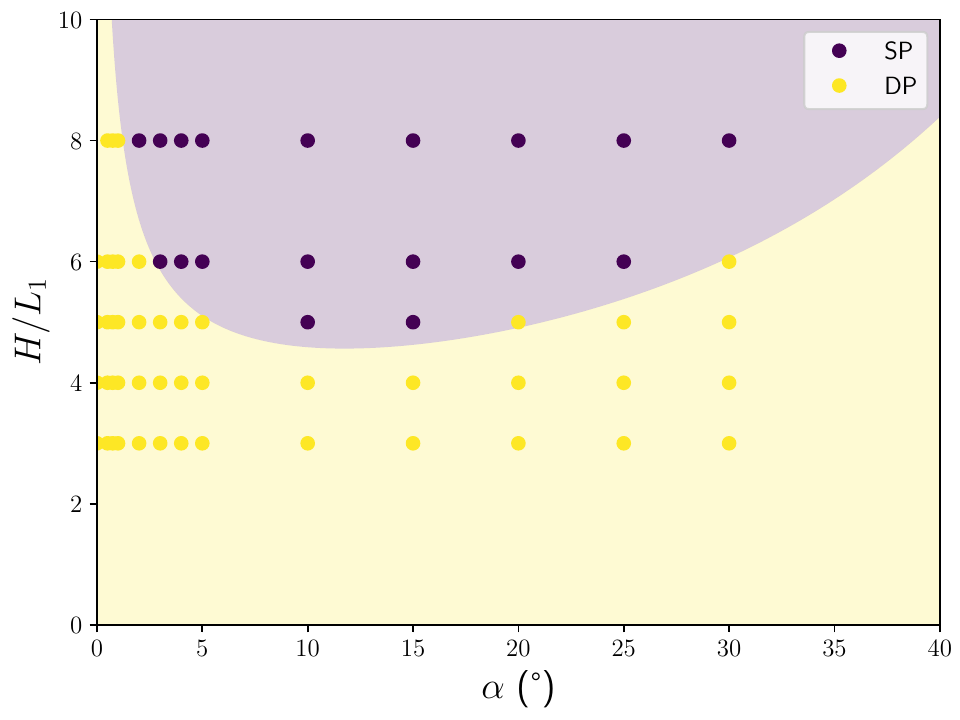}
	
	\smallskip
	(a) $k_BT/\varepsilon=1.1$
	
	\medskip
	
	\includegraphics[width=\columnwidth]{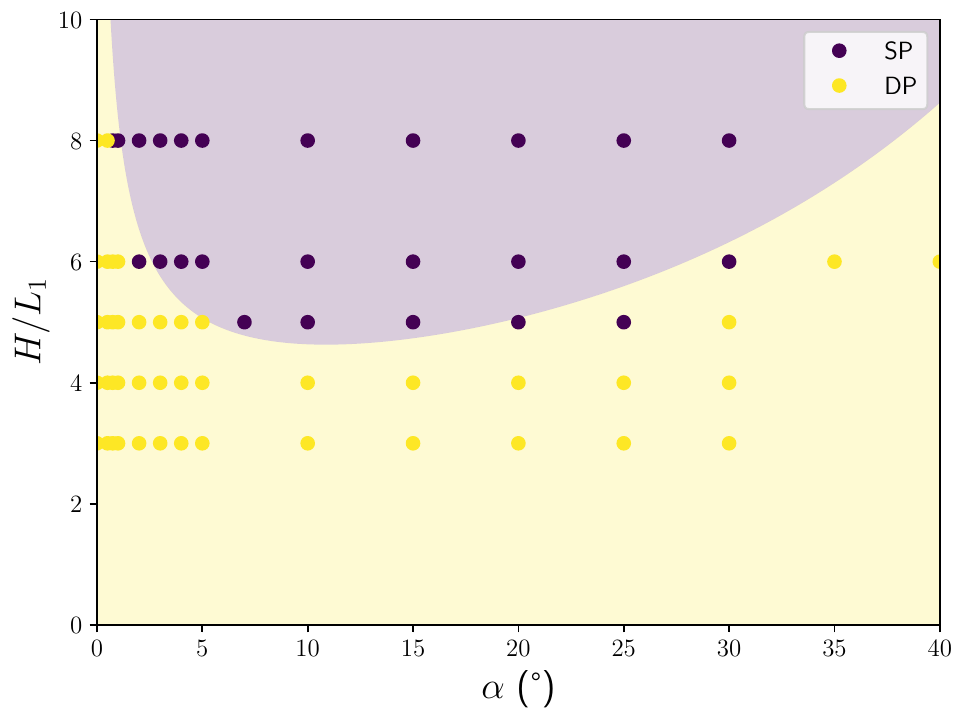}
	
	\smallskip
	(b) $k_BT/\varepsilon=1.2$
	
	\caption{Condensation morphologies in the symmetric open wedge in the $(\alpha,H/L_1)$ plane. Symbols show the DFT classification into single-pinned (SP) and double-pinned (DP) states, while the shaded regions and solid boundary represent the predictions of the macroscopic theory. Results are shown for (a) partially wet walls at $k_BT/\varepsilon=1.1$, with $\theta\simeq20^\circ$, and (b) completely wet walls at $k_BT/\varepsilon=1.2$, with $\theta=0$. The DFT results reproduce the nonmonotonic SP--DP boundary and the associated DP--SP--DP reentrant sequence. For completely wet walls, the SP region is systematically enlarged relative to the sharp-interface prediction.}
	\label{fig:sym_phase_diagram}
\end{figure}

\begin{figure*}
	\hfill
	\subfloat[$\alpha = 0 \degree$, DP]{
		\includegraphics[height=0.18\textheight]{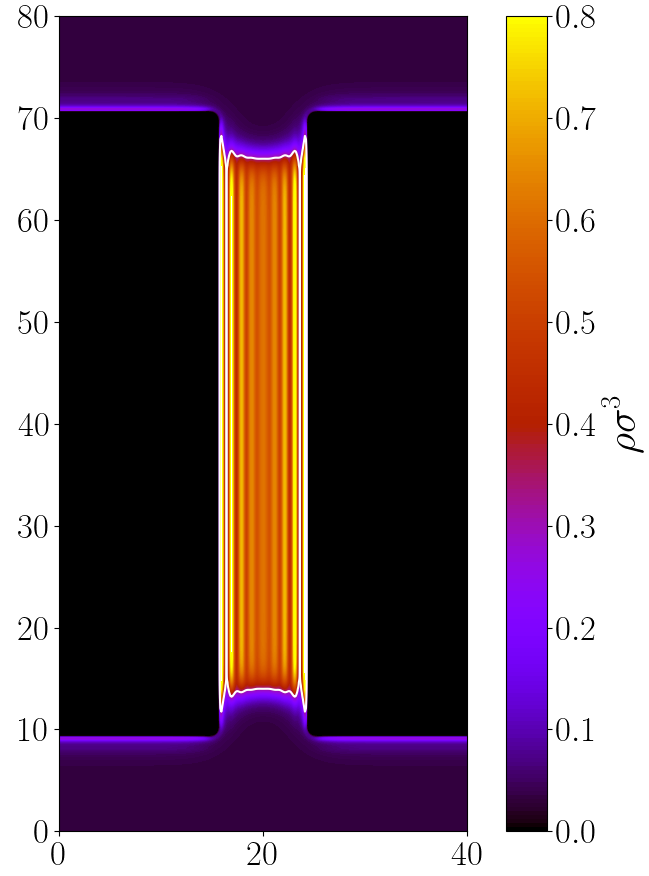}
		\label{subfig:symH6-alpha0-T1_1}
	}
	\hfill
	\subfloat[$\alpha = 2 \degree$, SP]{
		\includegraphics[height=0.18\textheight]{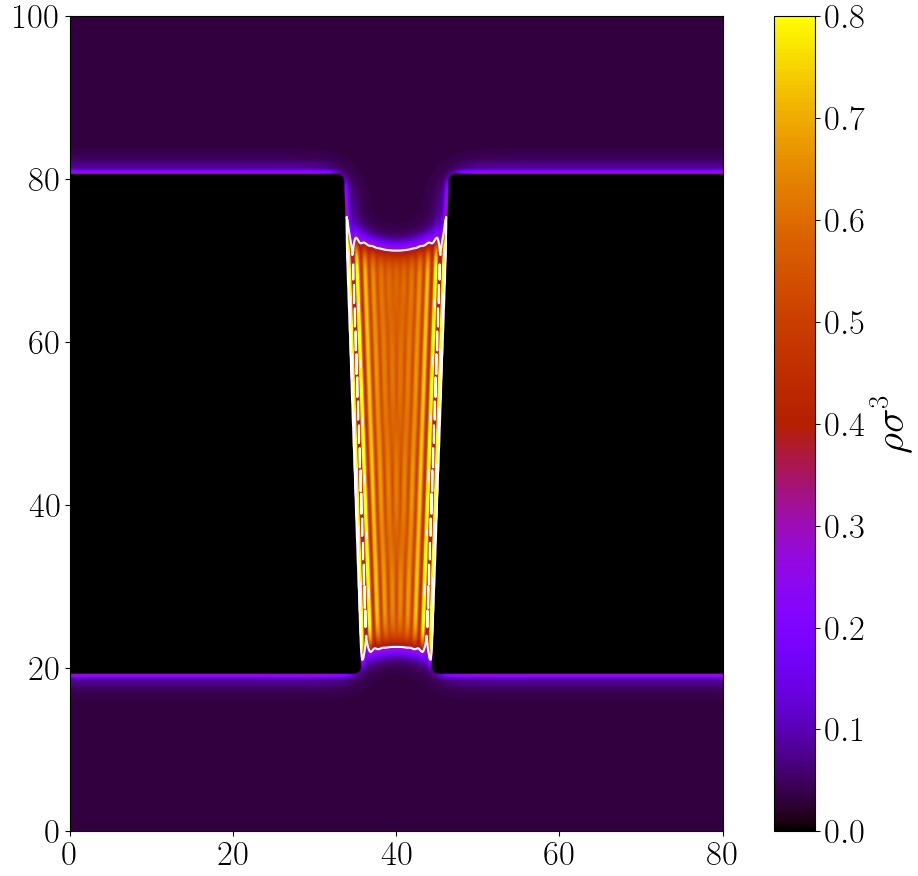}
		\label{subfig:symH6-alpha2-T1_1}
	}\hfill
	\subfloat[$\alpha = 10 \degree$, SP]{
		\includegraphics[height=0.18\textheight]{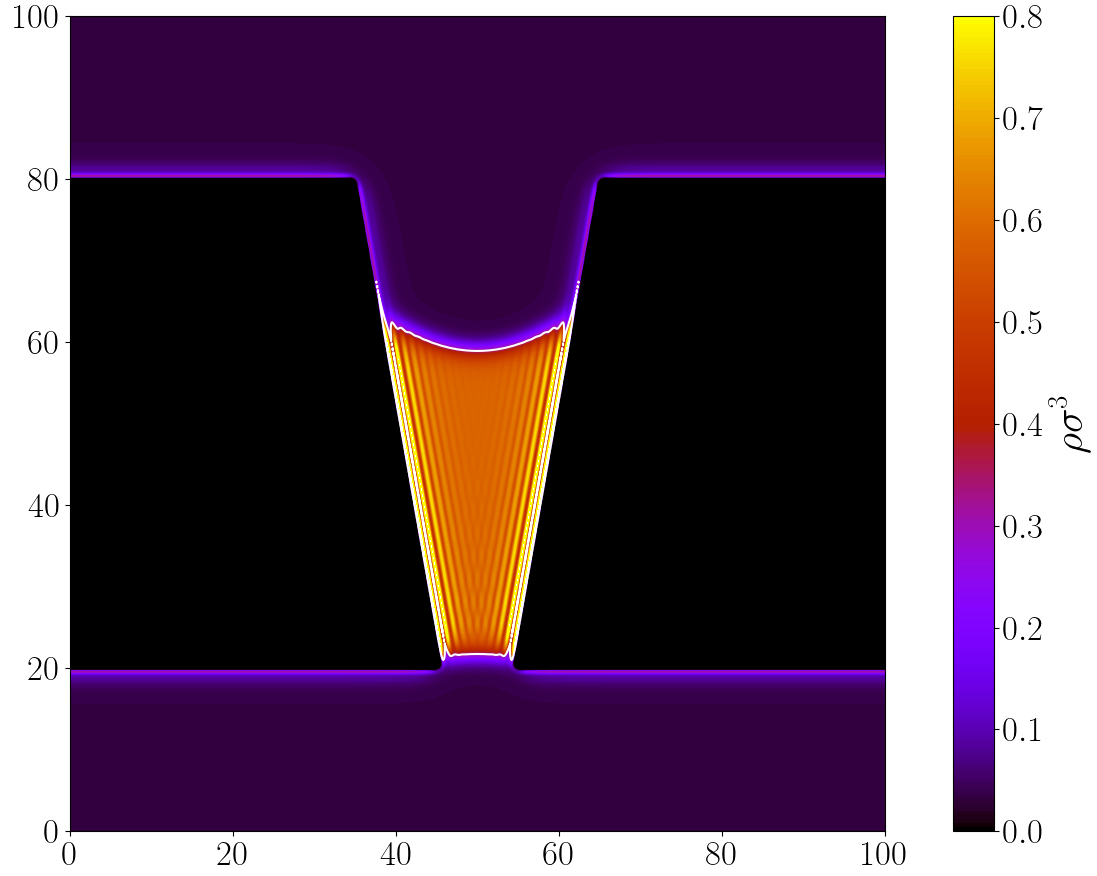}
		\label{subfig:symH6-alpha30-T1_1}
	}\phantom{}\\	
	\hfill\subfloat[$\alpha = 15 \degree$, SP]{
		\includegraphics[height=0.18\textheight]{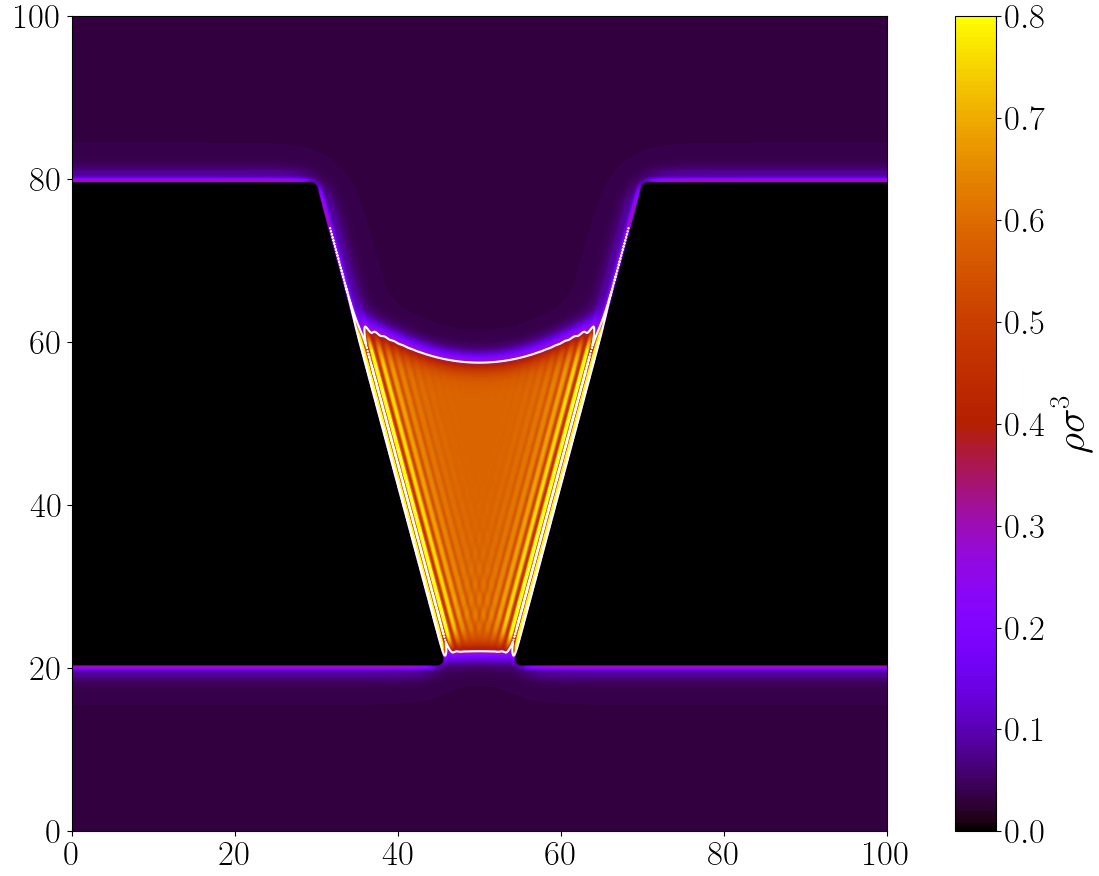}
		\label{subfig:symH6-alpha15-T1_1}
	}
	\hfill
	\subfloat[$\alpha = 25 \degree$, SP]{
		\includegraphics[height=0.18\textheight]{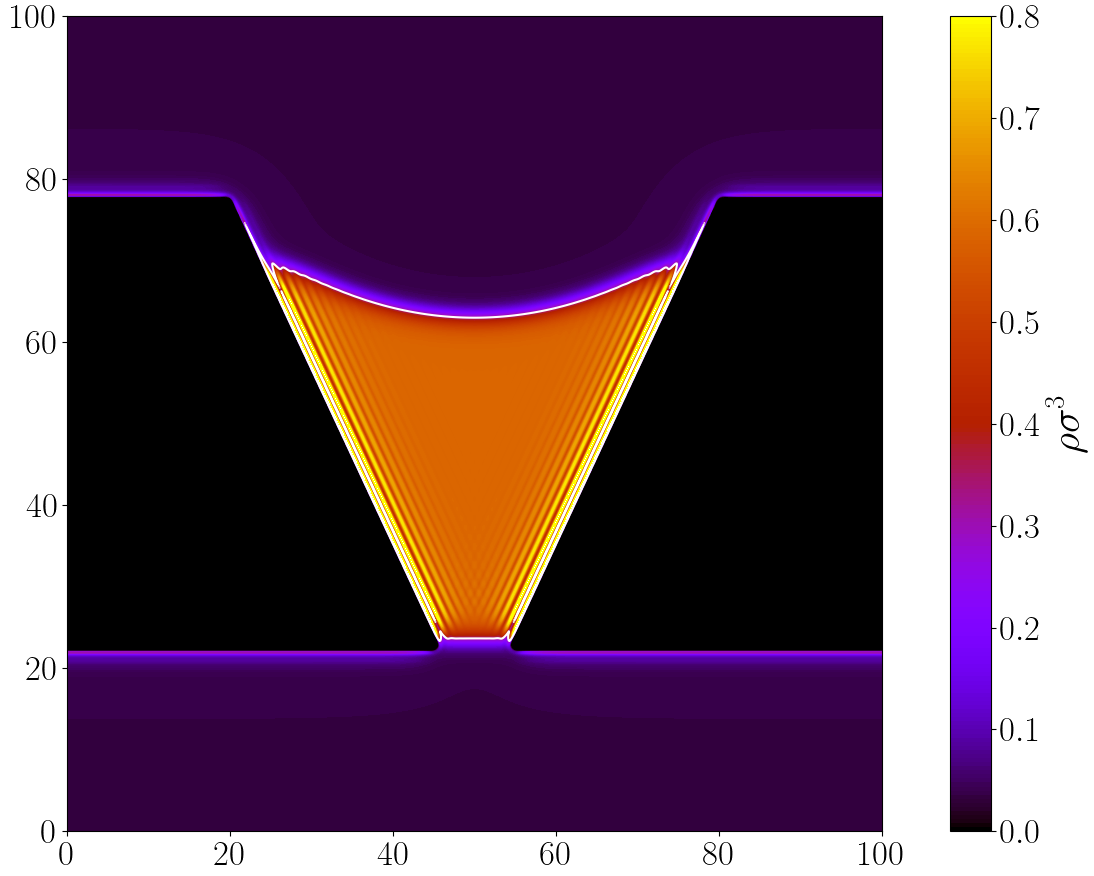}
		\label{subfig:symH6-alpha25-T1_1}
	}\hfill
	\subfloat[$\alpha = 30 \degree$, DP]{
		\includegraphics[height=0.18\textheight]{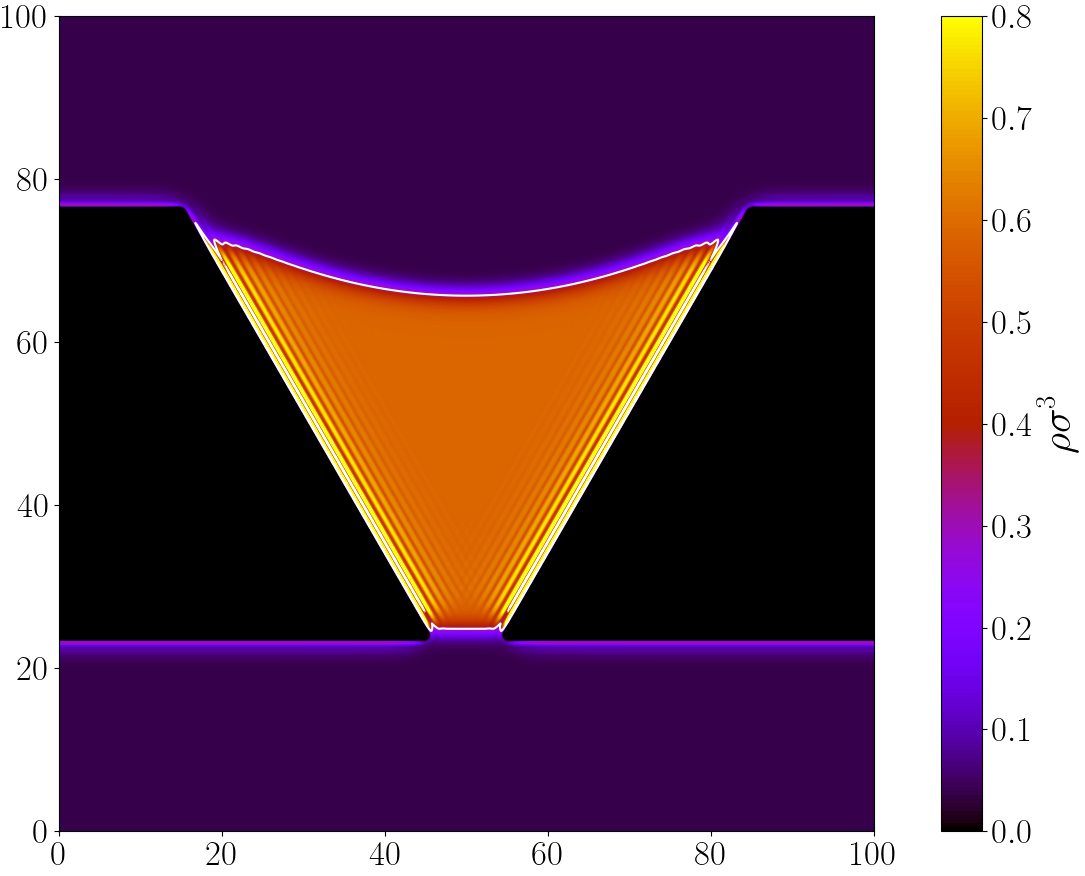}
		\label{subfig:symH6-alpha30-T1_1}
	}\hfill
	\phantom{}
	\caption{Representative equilibrium density profiles for the symmetric open wedge with $H/L_1=6$ at $k_BT/\varepsilon=1.1$. The wall inclination increases from (a) to (f), illustrating the reentrant sequence from double-pinned condensation at $\alpha=0^\circ$ to single-pinned condensation at intermediate inclinations and back to double-pinned condensation at $\alpha=30^\circ$. In the SP states, the wide-end meniscus lies inside the wedge and meets both walls at Young's contact angle, whereas in the DP states both menisci are pinned at the wedge openings.}
	\label{fig:sym_profiles}
\end{figure*}

\begin{figure}[tb]
	\centering
	\includegraphics[width=\columnwidth]{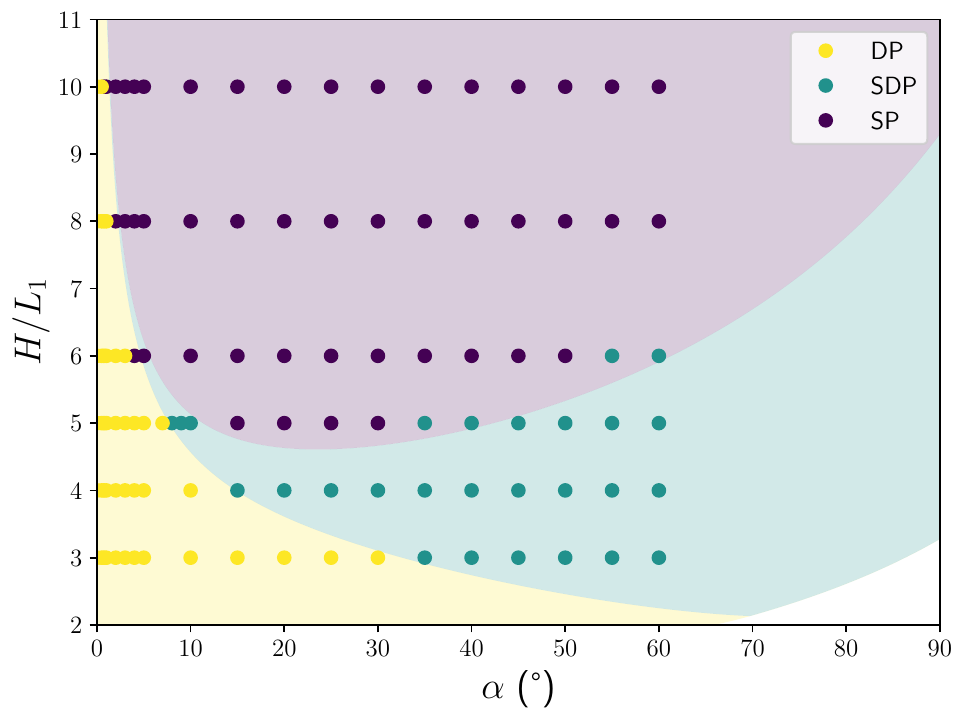}
	
	\smallskip
	(a) $k_BT/\varepsilon=1.1$
	
	\medskip
	
	\includegraphics[width=\columnwidth]{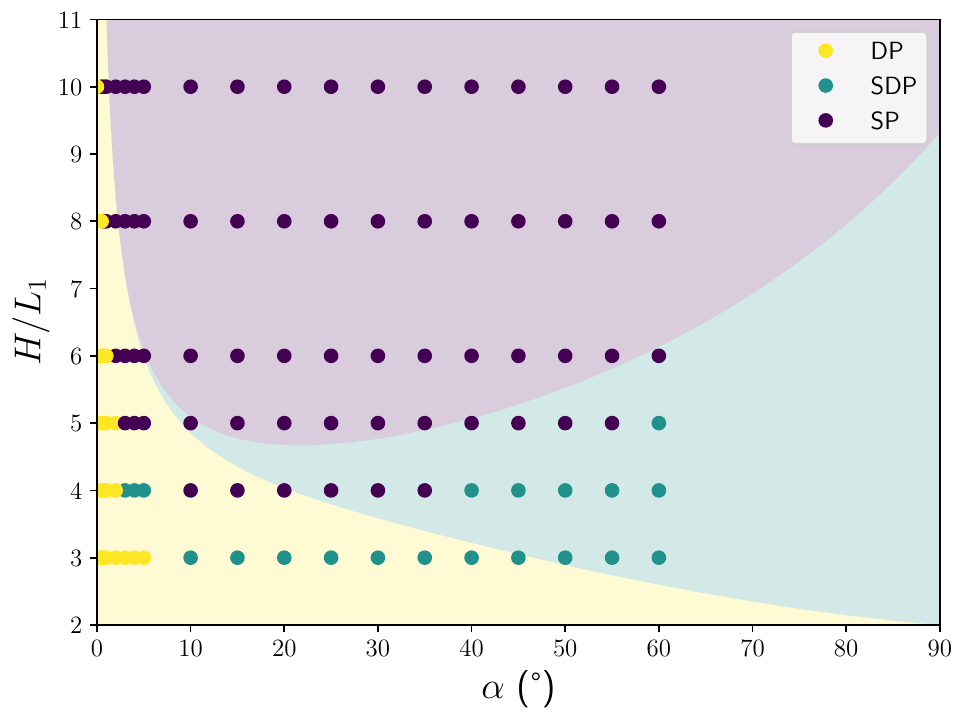}
	
	\smallskip
	(b) $k_BT/\varepsilon=1.2$
	
	\caption{Condensation morphologies in the asymmetric open wedge in the $(\alpha,H/L_1)$ plane. Symbols show the DFT classification into double-pinned (DP), semi-double-pinned (SDP), and single-pinned (SP) states, while the shaded regions and phase boundaries represent the macroscopic predictions. Results are shown for (a) partially wet walls at $k_BT/\varepsilon=1.1$, with $\theta\simeq20^\circ$, and (b) completely wet walls at $k_BT/\varepsilon=1.2$, with $\theta=0$. The DFT results reproduce the topology of the macroscopic phase diagrams, including the intermediate SDP region and the reentrant changes of morphology. The SP region is enlarged under complete-wetting conditions because of the liquid films adsorbed on the walls.}
	\label{fig:asym_phase_diagram}
\end{figure}

\begin{figure}[tb]
	\centering
	\includegraphics[width=\columnwidth]{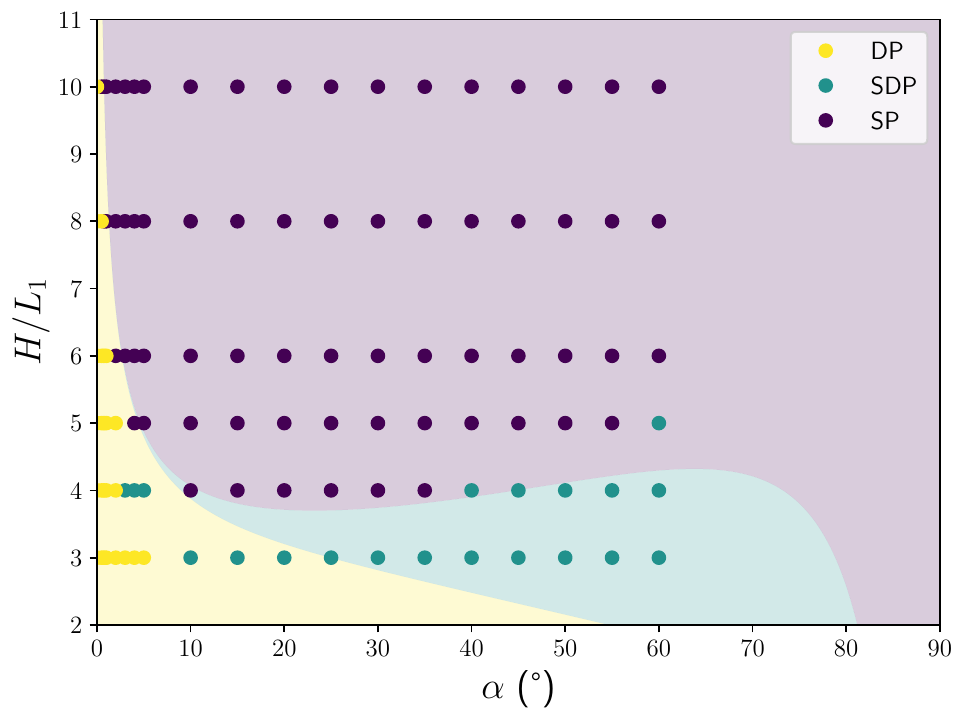}	
\caption{Condensation morphologies in the asymmetric open wedge under complete-wetting conditions, $k_BT/\varepsilon=1.2$, using the effective-width correction associated with the adsorbed liquid films. Symbols show the DFT classification into double-pinned (DP), semi-double-pinned (SDP), and single-pinned (SP) states, while the shaded regions show the macroscopic prediction evaluated using the effective narrow opening $L_{1,\mathrm{eff}}$. Compared with the original prediction in Fig.~\ref{fig:asym_phase_diagram}(b), the effective-width correction substantially improves the agreement with the DFT phase boundaries.}\label{fig:effective_width}
\end{figure}

\begin{figure*}
	\centering
	\begin{minipage}[t]{0.31\textwidth}
		\centering
		\includegraphics[width=\linewidth]{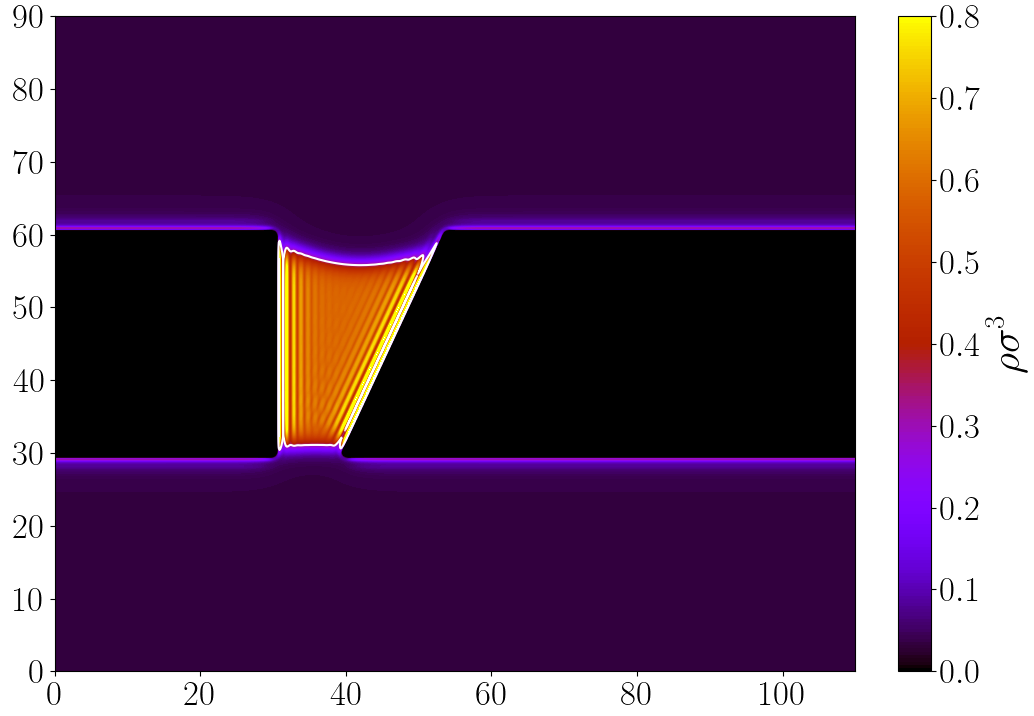}
		
		\smallskip
		(a) $H/L_1=3$, DP
	\end{minipage}
	\hfill
	\begin{minipage}[t]{0.31\textwidth}
		\centering
		\includegraphics[width=\linewidth]{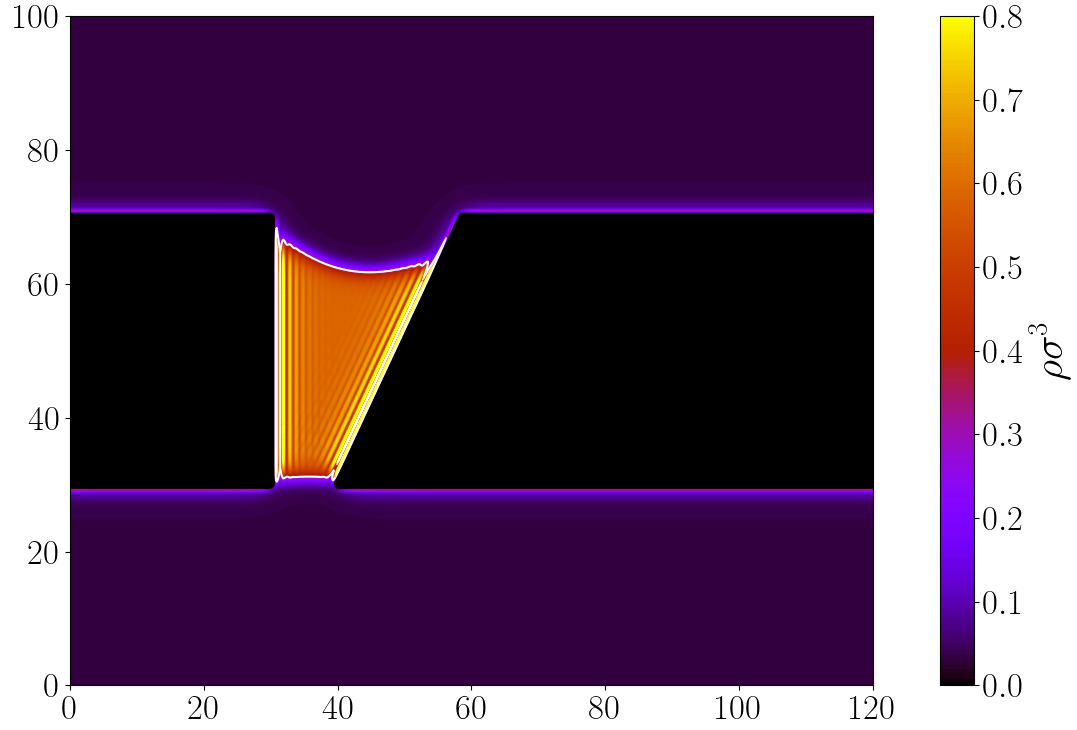}
		
		\smallskip
		(b) $H/L_1=4$, SDP
	\end{minipage}
	\hfill
	\begin{minipage}[t]{0.31\textwidth}
		\centering
		\includegraphics[width=\linewidth]{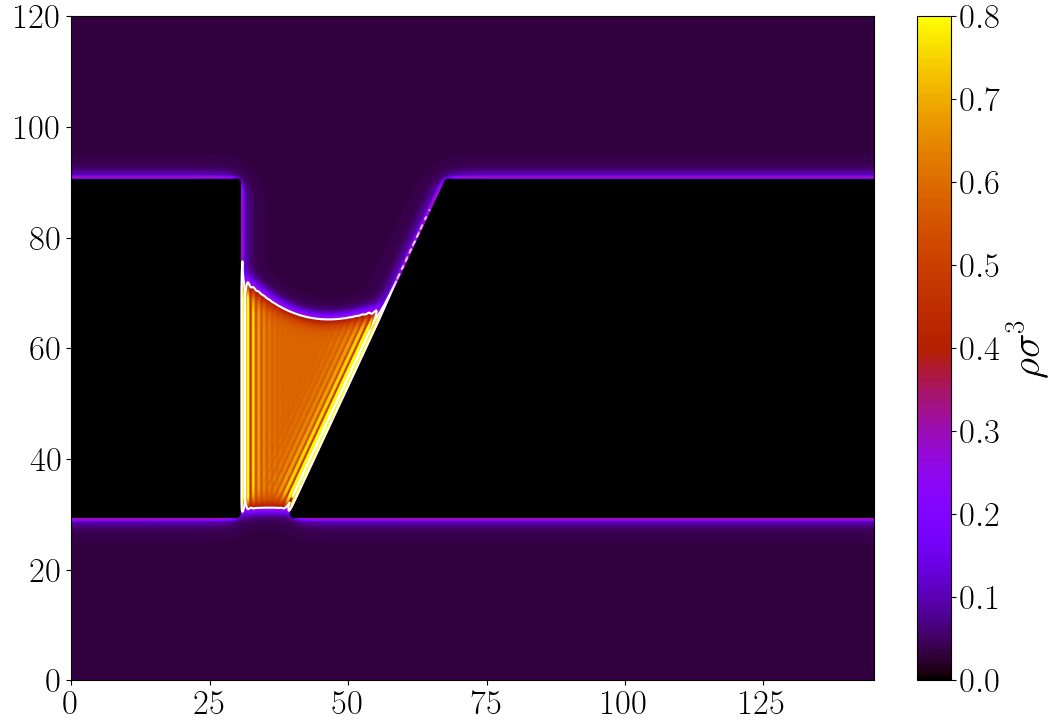}
		
		\smallskip
		(c) $H/L_1=6$, SP
	\end{minipage}
	
	\caption{Representative equilibrium density profiles for the asymmetric open wedge at $k_BT/\varepsilon=1.1$ and fixed opening angle $\alpha=25^\circ$. Increasing the wall length changes the condensed state from (a) DP at $H/L_1=3$, through (b) SDP at $H/L_1=4$, to (c) SP at $H/L_1=6$. The sequence provides a direct microscopic illustration of the three condensation morphologies predicted by the macroscopic theory. In the DP state, the upper meniscus is pinned at both edges of the wide opening; in the SDP state, it is pinned at the upper edge of the vertical wall and meets the inclined wall at Young's angle; in the SP state, it lies inside the wedge and meets both walls at Young's angle.}
	\label{fig:asym_profiles_H}
\end{figure*}

\begin{figure*}
	\centering
	\begin{minipage}[t]{0.31\textwidth}
		\centering
		\includegraphics[width=\linewidth]{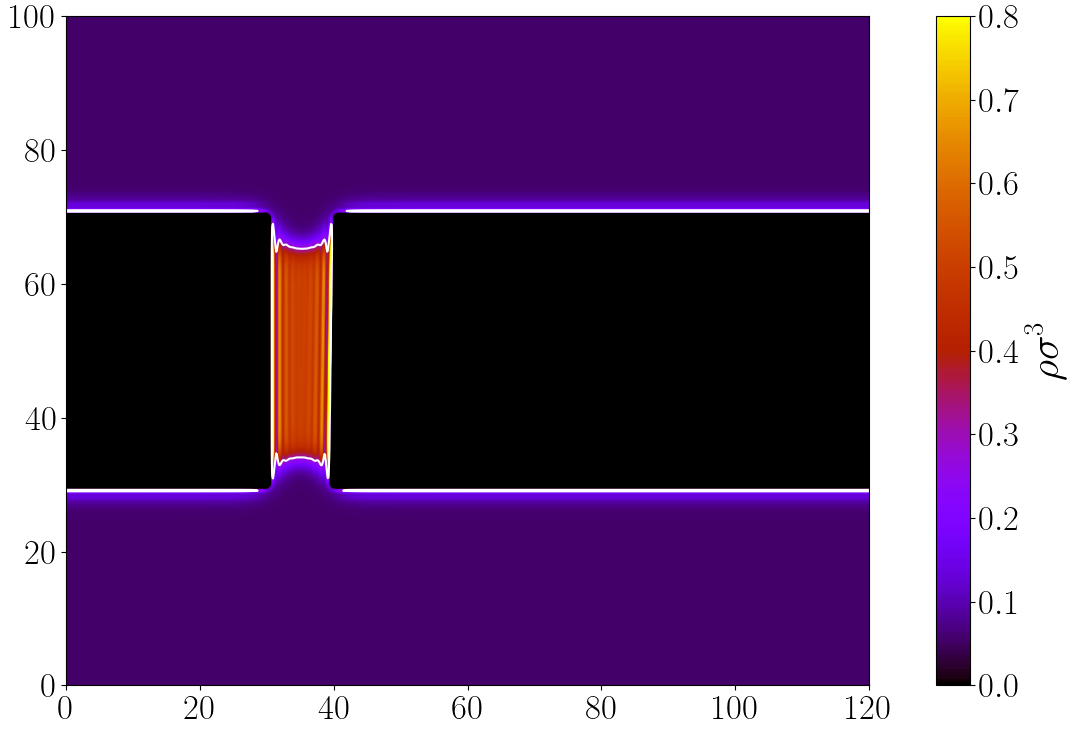}
		
		\smallskip
		(a) $\alpha=0.75^\circ$, DP
	\end{minipage}
	\hfill
	\begin{minipage}[t]{0.31\textwidth}
		\centering
		\includegraphics[width=\linewidth]{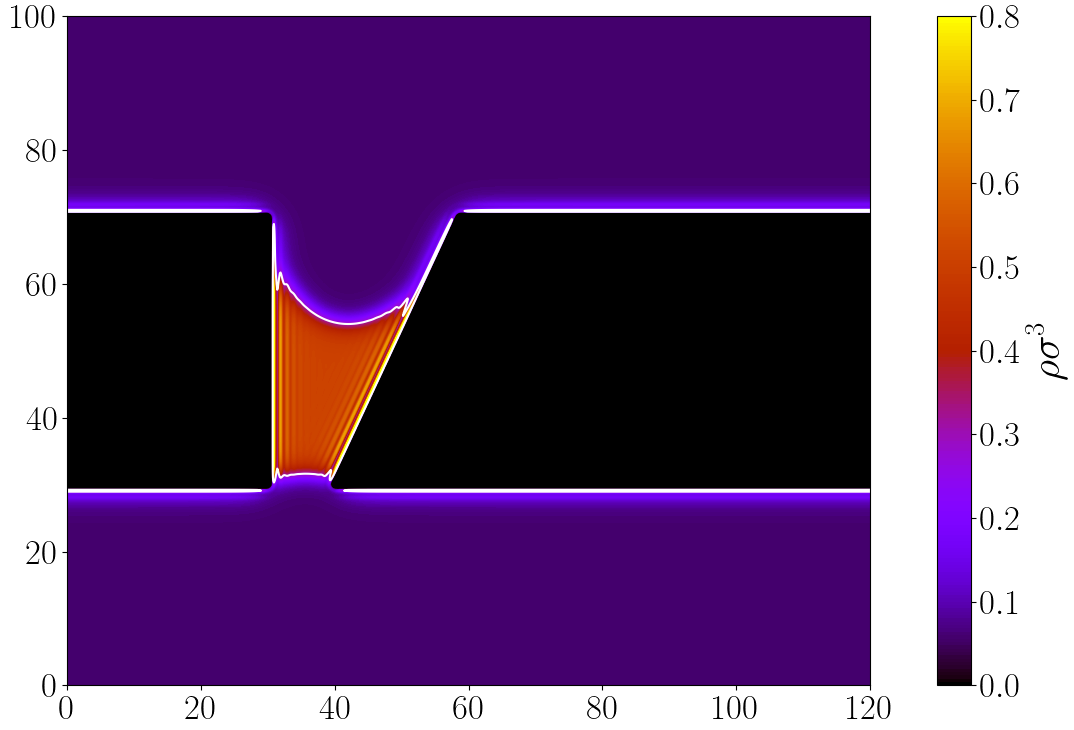}
		
		\smallskip
		(b) $\alpha=25^\circ$, SP
	\end{minipage}
	\hfill
	\begin{minipage}[t]{0.31\textwidth}
		\centering
		\includegraphics[width=\linewidth]{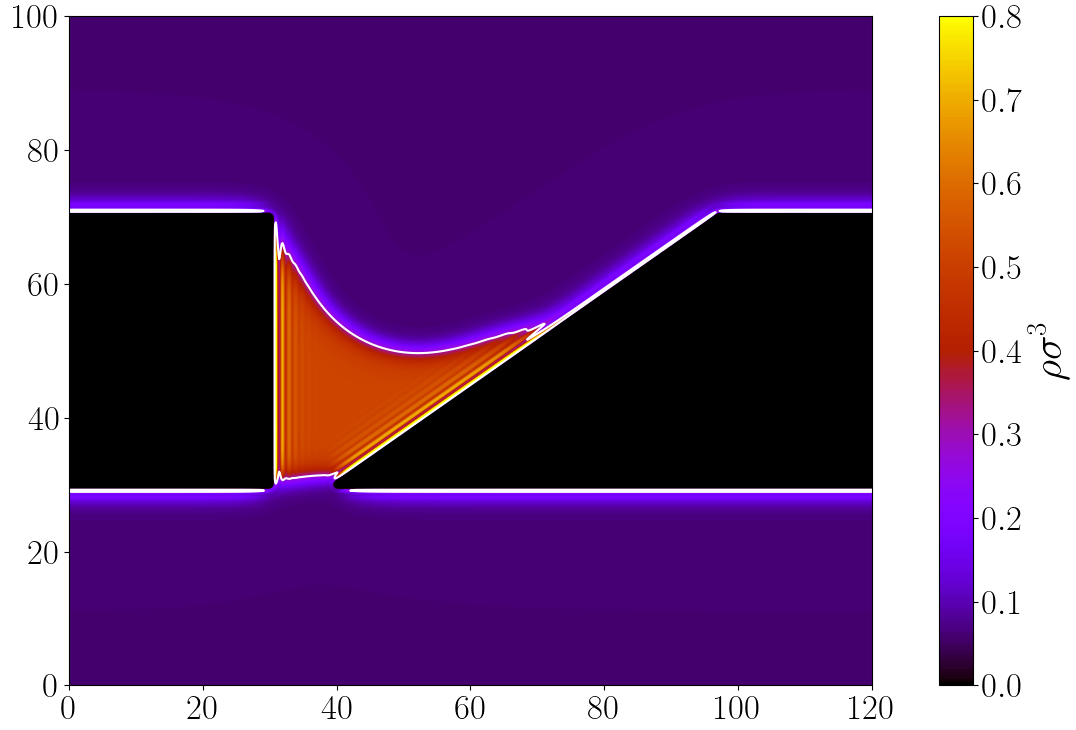}
		
		\smallskip
		(c) $\alpha=55^\circ$, SDP
	\end{minipage}
	
	\caption{Representative equilibrium density profiles for the asymmetric open wedge at $k_BT/\varepsilon=1.2$ and fixed wall length $H/L_1=4$. Increasing the opening angle changes the condensed state from (a) DP at $\alpha=0.75^\circ$, through (b) SP at $\alpha=25^\circ$, to (c) SDP at $\alpha=55^\circ$.}	
	\label{fig:asym_profiles_alpha}
\end{figure*}

For both the symmetric and asymmetric open wedges, the external potential is generated by walls that are assumed to be uniform and infinitely extended in the direction perpendicular to the two-dimensional wedge cross section. The system is therefore translationally invariant along the $y$ axis, and the resulting wall field depends only on $x$ and $z$. Explicit analytical expressions for the external potentials corresponding to both wedge geometries are given in the \SM{}~\cite{SupplementalMaterial}.\nocite{Malijevsky2014,Malijevsky2018} Their numerical implementation is also available as a Fortran module on GitHub~\cite{github}.

\begin{table}
	\label{tab:fluid-prop}
\caption{Bulk and interfacial properties of the model fluid at the two temperatures considered. Here $\rho_l$ and $\rho_g$ denote the coexisting liquid and gas densities, respectively, $\gamma$ is the liquid--gas surface tension, and $\theta$ is Young's contact angle at the corresponding planar wall.}	
	\centering
	\begin{tabular}{ccccccc}
		\toprule
		$\kB T/\varepsilon$ & $\rho_l \sigma^3$ & $\rho_g \sigma^3$ & $\gamma \sigma^2/\varepsilon$  & $\theta$ \\
		\midrule
		1.1 & 0.590 & 0.037 & 0.282 & 20\textdegree \\
		1.2 & 0.518 & 0.062 & 0.154 &  0 \\
		\bottomrule   
	\end{tabular}
\end{table}

The bulk critical temperature of the model is $k_BT_c/\varepsilon\simeq1.41$. For the chosen wall parameters, the corresponding planar wall undergoes a wetting transition at $k_BT_w/\varepsilon=1.12$. For both wedge geometries, we consider two subcritical temperatures: $k_BT/\varepsilon=1.1$, in which case the walls are only partially wet with Young's contact angle $\theta\simeq20^\circ$, and  $k_BT/\varepsilon=1.2$, which is above the wetting temperature, so that the walls are completely wet ($\theta=0$). Comparing these two distinct adsorption regimes allows us to assess the influence of the adsorbed liquid films that develop under complete-wetting conditions but are absent in the macroscopic description. The corresponding bulk and interfacial properties are listed in Table~I.

Minimization of Eq.~\eqref{eq:grand_potential} yields the self-consistent Euler-Lagrange equation
\begin{equation}
	\rho(\mathbf r)=\Lambda^{-3}\exp\!\left[\beta\mu-\beta V(\mathbf r)+c^{(1)}(\mathbf r)\right],
	\label{eq:euler_lagrange}
\end{equation}
where $c^{(1)}(\mathbf r)=c_{\rm hs}^{(1)}(\mathbf r)+c_{\rm att}^{(1)}(\mathbf r)$ is the one-body direct correlation function. The hard-sphere contribution is
\begin{equation}
	c_{\rm hs}^{(1)}(\mathbf r)
	=-\sum_\nu\int d\mathbf r'\,
	\frac{\partial\Phi(\{n_\nu(\mathbf r')\})}{\partial n_\nu}
	\omega_\nu(\mathbf r'-\mathbf r),
	\label{eq:chs}
\end{equation}
where the sum extends over the scalar and vector weighted densities, with the appropriate scalar product understood for the vector terms. The attractive contribution is
\begin{equation}
	c_{\rm att}^{(1)}(\mathbf r)
	=-\beta\int d\mathbf r'\,
	u_{\rm att}(|\mathbf r-\mathbf r'|)\rho(\mathbf r').
	\label{eq:catt}
\end{equation}

The Euler–Lagrange equation is solved on a uniform two-dimensional grid with spacing $0.1\sigma$. Gas-like and condensed solutions are obtained from appropriate initial profiles, and the condensation point is determined from equality of their grand potentials. Further numerical details are given in Ref.~\cite{Pospisil2022}.

In all calculations, the narrow opening is fixed at $L_1=10\sigma$, with $3\leq H/L_1\leq10$. The inclination angle is varied up to $40^\circ$ for the symmetric wedge and $60^\circ$ for the asymmetric wedge. The large number of parameter combinations is managed using the \texttt{signac} and \texttt{signac-flow} workflow packages \cite{Adorf2018,Ramasubramani2018}. The classification of the condensed states is described in the Supplemental Material.
	
\section{Numerical results}
\label{sec:results}

We now compare the microscopic DFT results with the macroscopic predictions summarized in Sec.~II. For each geometry, we construct phase diagrams in the $(\alpha,H/L_1)$ plane from the equilibrium condensed states obtained at $k_BT/\varepsilon=1.1$ and $1.2$, corresponding to partial and complete wetting, respectively. The DFT classifications are compared directly with the macroscopic phase boundaries, allowing us to assess both the predicted topology of the phase diagrams and the quantitative accuracy of the macroscopic theory.

\subsection{Symmetric open wedge}
\label{subsec:results_symmetric}

The phase diagrams for the symmetric wedge are shown in Fig.~\ref{fig:sym_phase_diagram}. For partially wet walls, the DFT results follow the macroscopic SP-DP boundary remarkably closely. For the shortest wedges, approximately $H/L_1\lesssim5$, condensation remains of DP type throughout the range of opening angles, in close agreement with the macroscopic prediction. As $H/L_1$ is increased, the SP region first appears near $\alpha\simeq10^\circ$, again in very good agreement with the theory.  The resulting nonmonotonic boundary gives rise to the reentrant sequence DP $\to$ SP $\to$ DP upon increasing $\alpha$ at fixed $H/L_1$, as predicted.

For completely wet walls, the topology of the DFT phase diagram remains the same, but the SP region is slightly wider than predicted by the macroscopic theory. In particular, SP condensation occurs for somewhat shorter wedges and over a broader range of inclinations. This displacement is associated with the liquid films adsorbed on the completely wet walls, an aspect absent from the macroscopic theory. The films reduce the effective width of the gas-filled region and thereby increase the effective aspect ratio of the confinement, favouring the SP morphology.

Representative equilibrium density profiles for $H/L_1=6$ and $k_BT/\varepsilon=1.1$ are shown in Fig.~\ref{fig:sym_profiles}. At $\alpha=0$, the system exhibits the usual DP condensation of a finite parallel slit. However, already at a very small nonzero inclination, the wide-end meniscus detaches from the opening and moves inside the wedge, producing the SP state. The SP morphology persists over a broad interval of intermediate inclinations. At sufficiently large $\alpha$, the meniscus forms at the wide opening again and the DP configuration is restored. The profiles thus provide a direct microscopic illustration of the predicted DP--SP--DP reentrance.

\subsection{Asymmetric open wedge}
\label{subsec:results_asymmetric}

The asymmetric geometry exhibits a richer phase behavior because the upper meniscus may be free, pinned at one edge, or pinned at both edges. The resulting phase diagrams are shown in Fig.~\ref{fig:asym_phase_diagram}. The DFT results reproduce the three regions predicted by the macroscopic theory: DP region at small opening angles and for relatively short wedges, SDP region at moderate and large opening angles for shorter wedges, and SP region at larger values of $H/L_1$.

For partially wet walls, the agreement with the macroscopic phase boundaries is quantitatively accurate. Depending on the wedge aspect ratio, increasing $\alpha$ produces several possible sequences. The shortest wedges undergo only a DP $\to$ SDP change, whereas somewhat longer wedges exhibit the sequence DP $\to$ SP $\to$ SDP. Within a narrow interval of wedge sizes near  $H/L_1=5$, the more complex reentrant sequence DP $\to$ SDP $\to$ SP $\to$ SDP occurs. For sufficiently long wedges, the SP state remains stable throughout the entire range of inclinations considered.


For completely wet walls, the qualitative topology predicted by the macroscopic theory is again retained, but the SP region expands at the expense of both the DP and SDP regions. As for the symmetric wedge, the shift is associated with liquid films adsorbed on the walls. These films reduce the effective width available to the gas and thereby increase the effective aspect ratio, favouring equilibrium states characteristic of longer wedges.  To assess this interpretation more quantitatively, we estimated the thickness $\ell_\pi$ of the adsorbed liquid layer from the DFT density profiles and used it to define an effective narrow opening. For the present estimate, we use $\ell_\pi/L_1 \simeq 0.1$
($\ell_\pi\simeq\sigma$ for $L_1=10\sigma$), as estimated from
the DFT density profiles at condensation. This approach is motivated by the Derjaguin correction to Kelvin-like descriptions of capillary condensation, in which wetting layers effectively reduce the space available to the gas phase~\cite{mal26}. For the symmetric wedge, we define
	\begin{equation}
		L_{1,\mathrm{eff}}=L_1-2\ell_\pi\sec\alpha,
	\end{equation}
	whereas for the asymmetric wedge
	\begin{equation}
		L_{1,\mathrm{eff}}=L_1-\ell_\pi(1+\sec\alpha).
	\end{equation}
	The corresponding effective aspect ratio,
	\begin{equation}
		\left(\frac{H}{L_1}\right)_{\mathrm{eff}}
		=\frac{H}{L_{1,\mathrm{eff}}},
	\end{equation}
	is therefore larger than the bare value. Using this effective aspect ratio in the macroscopic theory shifts the predicted phase boundaries in the same direction as the DFT results and accounts for a significant part of the observed difference. This is demonstrated in Fig.~\ref{fig:effective_width} for the asymmetric wedge, where the phase diagram obtained using $L_{1,\mathrm{eff}}$ shows a markedly improved agreement with the microscopic classification. In particular, the corrected boundaries reproduce much more closely the extent of the SP region and its boundaries with the DP and SDP states. This supports the interpretation that the systematic deviations under complete-wetting conditions arise predominantly from the finite thickness of the adsorbed liquid films. 

Representative equilibrium density profiles at $k_BT/\varepsilon=1.1$ and fixed $\alpha=25^\circ$ are shown in Fig.~\ref{fig:asym_profiles_H}. Increasing $H/L_1$ from 3 to 6 drives the sequence DP $\to$ SDP $\to$ SP. In the DP state, the upper meniscus spans the wide opening and is pinned at both edges. In the SDP state, it remains pinned at the upper edge of the vertical wall but meets the inclined wall at Young's angle. In the SP state, it lies entirely inside the wedge and meets both walls at the equilibrium contact angle.

A complementary sequence is illustrated in Fig.~\ref{fig:asym_profiles_alpha} for $k_BT/\varepsilon=1.2$ and fixed wall length $H/L_1=4$. As $\alpha$ increases, the condensation states change from DP to SP and finally to SDP.  The profiles also illustrate the pronounced adsorption films formed under complete-wetting conditions.

\section{Conclusion}
\label{sec:conclusions}

We have used classical density functional theory to test the macroscopic description of first-order condensation in finite open wedges. Two geometries were considered: a symmetric wedge formed by two identical, equally inclined walls and an asymmetric wedge formed by a vertical and an inclined wall of equal vertical extent. In both cases, calculations were performed at two temperatures corresponding to partially wet walls, $T<T_w$, and completely wet walls, $T>T_w$.

For the symmetric wedge, the DFT calculations confirm the existence of the single-pinned (SP) and double-pinned (DP) condensed states and reproduce the non-monotonic phase boundary separating them. In particular, the predicted reentrant sequence DP $\to$ SP $\to$ DP is observed upon increasing the wall inclination at fixed wedge length. For the asymmetric wedge, the microscopic calculations also confirm the additional semi-double-pinned (SDP) state arising from partial depinning of the wide-end meniscus. The resulting phase diagrams reproduce the predicted DP, SDP, and SP regions, including the reentrant changes of morphology with wedge size and opening angle.

For partially wet walls, the agreement between the microscopic calculations and the macroscopic theory is nearly perfect, despite the relatively small narrow opening, $L_1=10\sigma$, used in the DFT calculations. This shows that the macroscopic construction captures both the possible condensation morphologies and their stability boundaries down to molecular length scales.

For completely wet walls, the topology of the phase diagrams remains unchanged, but the SP region is systematically enlarged relative to the macroscopic prediction. This deviation has the same origin in both geometries. The adsorbed liquid films reduce the effective width available to the gas phase and thereby increase the effective aspect ratio of the wedge,  favouring morphologies characteristic of longer wedges. 

Overall, the results show that the macroscopic theories provide a reliable and accurate description of first-order condensation in finite open wedges and that the concepts of edge pinning and meniscus depinning remain meaningful even at molecular scales. A natural next step is to develop a mesoscopic theory in which adsorption films enter explicitly, for example through a binding potential or an effective wedge geometry. Such a theory could provide quantitative corrections to the phase boundaries under complete-wetting conditions and predict their systematic dependence on the geometric parameters and the range of the wall-fluid forces. It would also be interesting to determine whether the same sequence of pinned morphologies survives for chemically heterogeneous walls, rounded edges, or three-dimensional wedge geometries, where additional line and edge contributions may become relevant.

%

\end{document}